\documentclass[prd,nofootinbib,tightenlines]{revtex4}
\usepackage{graphicx}
\def\journal#1, #2, #3#4, #5#6#7#8    {
    {#1~} {#2}  (#5#6#7#8) #3#4}

\usepackage{amssymb}
\usepackage{amsmath}
\usepackage{lscape}
\begin{document}


\renewcommand{\thesection}{\arabic{section}}
\renewcommand{\thesubsection}{\thesection.\arabic{subsection}}
\renewcommand{\theequation}{\arabic{equation}}
\renewcommand {\c}  {\'{c}}
\newcommand {\cc} {\v{c}}
\newcommand {\s}  {\v{s}}
\newcommand {\CC} {\v{C}}
\newcommand {\C}  {\'{C}}
\newcommand {\Z}  {\v{Z}}
\newcommand{\pv}[1]{{-  \hspace {-4.0mm} #1}}

\newcommand{\be}{\begin{equation}} \newcommand{\ee}{\end{equation}}
\newcommand{\bea}{\begin{eqnarray}}\newcommand{\eea}{\end{eqnarray}}
\newcommand{\grad}{\bm \nabla}

\baselineskip=14pt

{\small\hfill SINP/TNP/2009/12}
\begin{center}
{\bf  \Large  Scattering in graphene with impurities : A low energy effective theory} 
\bigskip
 
Kumar S. Gupta {\footnote{e-mail: kumars.gupta@saha.ac.in}}\\
Theory Division, Saha Institute of Nuclear Physics, 1/AF Bidhannagar, Calcutta 700064, India\\[3mm]

\bigskip

Andjelo Samsarov {\footnote{e-mail: asamsarov@irb.hr}} \\  
 Rudjer Bo\v{s}kovi\'c Institute, Bijeni\v cka  c.54, HR-10002 Zagreb,
Croatia \\[3mm]

\bigskip

Siddhartha Sen \footnote{e-mail: sen@maths.ucd.ie}\\
   School of Mathematical Sciences, UCD, Belfield, Dublin 4, Ireland\\
  Department of Theoretical Physics, Indian Association for the
   Cultivation of Science, Calcutta - 700032, India\\[3mm]

\end{center}
\setcounter{page}{1}
\bigskip


\begin{abstract}

We analyze the scattering sector of the Hamiltonians for both gapless and gapped graphene in the presence of a charge impurity using the 2D Dirac equation, which is applicable in the long wavelength limit. We show that for certain range of the system parameters, the combined effect of the short range interactions due to the charge impurity can be modelled using a single real parameter appearing in the boundary conditions. The phase shifts and the scattering matrix depend explicitly on this parameter. We argue that this parameter for graphene can be fixed empirically, through measurements of observables that depend on the scattering data.

\bigskip 
\noindent
PACS number(s): 03.65.Ge, 81.05.Uw 
\end{abstract}


\maketitle
 


\section{Introduction}

The experimental fabrication of monolayer graphene \cite{novo,geim}, a two dimensional system with a hexagonal honeycomb structure, has opened up an area of physics which is of wide theoretical and experimental interest. The low energy dynamics of gapless graphene is described by a 2D massless Dirac equation \cite{sem}. The effect of charge impurities in gapless graphene has been studied extensively, where bound states are predicted when the Coulomb charge exceeds a certain critical value \cite{castro, levi1, levi2, us1}. Below this critical value, the bound states do not form in massless graphene due to the Klein paradox. The generalization of this model where a Dirac mass term is included in the Hamiltonian is known to describe gapped graphene. The effect of charge impurities in gapped graphene have also been studied \cite{ho, novi, kot1, kot2, us2}, where bound states can form for arbitrarily small value of the Coulomb charge. 

 Apart from providing an axially symmetric Coulomb potential, the charge impurities may also induce other short-range or singular interactions, such as a delta function type potential. The detailed knowledge of such short range interactions in any given sample would be very hard to ascertain empirically. Moreover, the 2D Dirac description for graphene is valid only in the low energy (long distance) limit and it would not be practical to include short range interactions directly in such a Hamiltonian. One possible way to model the combined effect of such short range interactions on the long wavelength dynamics is through the choice of appropriate boundary conditions.  

Recently we have analyzed the effects of generalized boundary conditions, that may arise in the presence of a charge impurity, for the bound state sectors of gapless \cite{us1} and gapped {\cite{us2} graphene. In these papers, the choice of appropriate boundary conditions due to the presence of charge impurities in graphene was guided by the principle of self-adjointness of the corresponding Hamiltonian. This is a well known approach that provides a reliable description of bound states in systems with delta function type potentials \cite{jackiw} as well as in fermionic \cite{ld1,ld2,ld3,ld4,ld5} and anyonic systems \cite{an1,an2} and in molecular physics \cite{mol}. It has already been used to study certain topological defects in graphene \cite{topo1,topo2}. Apart from bound states, this method is also useful to study the effects of boundary conditions on scattering states, phase shifts and $S$ matrices in systems with short range interactions \cite{biru1,biru2,biru3,biru4}. 
 Application of this approach in the analysis of bound states in graphene with charge impurities \cite{us1,us2} introduced a new parameter which labels the boundary conditions. This self-adjoint extension parameter has to be fixed empirically, possibly through the scanning tunneling microscopy (STM) measurements of the local density of states (LDOS).

In this paper, we analyze the scattering states of gapless and gapped graphene within the framework of an effective low energy description in the presence of a charge impurity. Here again we model the combined effect of the short range interactions due to the charge impurity through the choice of boundary conditions, which is guided by the principle of self-adjointness. In the scattering sector, the phase shifts and the $S$ matrices depend explicitly on the boundary conditions, through the choice of the self-adjoint extension parameter. The knowledge of the phase shifts can be used to calculate physical properties of the system, such as resistivity \cite{kats}. In addition, the effect of this new class of boundary conditions can also appear in the density of states and Friedel sum rule. In our formulation, all these observable effects would depend on the self-adjoint extension parameter, which can be used to determine it empirically. For consistency, we also demonstrate that 
 the bound state energies calculated in \cite{us2} can be recovered from the poles of the scattering matrix as well. 

This paper is organized as follows. In Section 2 we discuss the effect of boundary conditions on the scattering sector of the massless graphene with a charge impurity. In Section 3 a similar analysis is carried out for the gapped graphene with a charge impurity. We conclude this paper in Section 4 with a discussion of the experimentally observable effects of our analysis.
 
\section{Scattering in gapless graphene with charge impurity}

In this Section we shall analyze the scattering states, phase shifts and the $S$ matrix for gapless graphene in the presence of a charge impurity. We begin by a brief review of the gapless graphene system when a charge impurity is present. Next we discuss the allowed class of boundary conditions for this system and end by doing the analysis for the scattering sector using the new boundary conditions. We show that the resulting $S$ matrix and the phase shifts depend explicitly on a parameter which labels the boundary conditions. We argue that this parameter cannot be determined from theoretical analysis but must be fixed empirically.  

\subsection{Gapless graphene with a charge impurity}

The 2D Dirac equation describing a Coulomb type charge impurity in a gapless graphene monolayer can be written as 
\begin{equation} \label{mles1}
  H \Psi = E \Psi,
\end{equation}
with the Hamiltonian $H$ being given by 
\begin{equation} \label{mles2}
  H = -i \hbar v_{F} ( {\sigma}_{1} {\partial}_{x} + {\sigma}_{2} {\partial}_{y} )
     - \frac{\alpha}{r} \equiv 
 \hbar v_{F} \left( \begin{array}{cc}
    0 & - i {\partial}_{x} - {\partial}_{y}  \\
 - i {\partial}_{x} + {\partial}_{y}    & 0 \\
\end{array} \right) - \frac{\alpha}{r},
\end{equation}
where $r$ is the radial coordinate on the two dimensional $x-y$ plane. In above, $v_{F}$ is the graphene Fermi velocity and $\alpha = \frac{Ze^2}{\kappa}$, where $Z$ denotes the impurity valence, $e$ is the unit charge and $\kappa$ is the effective dielectric constant. One can separate Eq.(\ref{mles1}) by
 assuming the solution of the form
\begin{equation} \label{mles3} 
 \Psi (r, \phi) = \left( \begin{array}{c}
 {F (r ) ~ \frac{1}{\sqrt{2 \pi}} e^{i(j - \frac{1}{2})\phi} } \\
  {G (r )~ \frac{1}{\sqrt{2 \pi}} e^{i(j + \frac{1}{2})\phi} } \\
\end{array} \right),
\end{equation}
where $\phi $ denotes the corresponding polar angle and $j$ is the half integral azimuthal quantum number. Consider the ansatz
\begin{equation} \label{mles4}
 F(r) = 
    e^{ikr} {r}^{\nu - \frac{1}{2}}
    (u(r) + v(r) ), 
\end{equation}
\begin{equation} \label{mles5}
 G(r) = 
    e^{ikr} {r}^{\nu - \frac{1}{2}}
    (u(r) - v(r) ),
\end{equation}
 where $\; \nu = \sqrt{{j}^2 - {\beta}^2}, \;\; \beta = \frac{\alpha}{
 \hbar v_{F} } \equiv \frac{Z
 e^2}{\kappa \hbar v_{F} }, \;\; k
 = - \frac{E}{\hbar v_{F}}. \; $  
This leads to the radial Dirac operator 
\begin{equation} \label{raddirac}
H_r = \left( \begin{array}{cc}
r \frac{d}{d r} + \nu + i \beta + 2ikr & -j  \\
- j  & r \frac{d}{d r} + \nu - i \beta \\
\end{array} \right)
\end{equation}
and the set of coupled equations
\begin{equation} \label{mles6}
   r \frac{d u}{d r} + (\nu + i \beta + 2ikr) u - j v = 0,
\end{equation}
\begin{equation} \label{mles7}
  r \frac{d v}{d r} + (\nu - i\beta ) v - j u = 0.
\end{equation}
  After introducing a
variable $\; z = -2ikr, \; $ the last two equations can be combined into a single 
equation for the function $v$, given by
\begin{equation} \label{mles7.1}
 z \frac{d^{2} v }{d z^{2}} + (1 + 2\nu - z)
 \frac{d v}{d z} - \left ( \nu - i \beta \right ) v = 0. 
\end{equation}
This shows that $v$ is described by a confluent hypergeometric function \cite{AS} and a similar conclusion holds for the function $u$ as well. 

In what follows, we shall set $\hbar = v_{F} = 1$.
It may be noted that since the minimum magnitude of the azimuthal quantum number is $\; j = \frac{1}{2}, \; $ the corresponding critical value of $\; \beta \; $ is given by $\; {\beta}_{crit} = \frac{1}{2}.\; $ In this paper, as far as the massless graphene case is considered,
we shall be interested only in the subcritical region. In the subcritical region, the parameter $\; \nu \;$ is a positive real number with $\; \nu = 0 \;$ denoting the critical value.

\subsection{Boundary conditions for gapless graphene with a charge impurity}

Equation (\ref{raddirac})  defines the radial Dirac operator $\; H_{r} \; $ for a gapless graphene with a charge impurity located at $\; r = 0. $ As discussed before, the impurity can give rise to additional short range interactions which cannot be incorporated in the Dirac description valid in the long wavelength limit. We can however model the combined effect of these short range interactions through suitable boundary conditions. We shall be guided in our search for the suitable boundary conditions by the requirement of self-adjointness of $\; H_{r} \; $, which is essential for the conservation of probabilities and for the unitary time evolution of the system. For gapless graphene, this has already been discussed in \cite{us1}, which we shall review and elaborate below.

The operator $\; H_{r} \; $ is symmetric (or Hermitian) \cite{reed} in the domain  $\; D(H_{r}) \; $ defined by
\begin{equation} \label{mles7.2}
  D(H_{r}) = \{ \Psi \; | \; \Psi (0)={\Psi}^{'} (0) =0, \;\; \Psi,
 {\Psi}^{'} \;
    \mbox{absolutely continuous}, \;\; \Psi \in L^2 (rdr)  \}.
\end{equation}
In order to see if the operator $\; H_{r} \; $ is already self-adjoint in $D(H_{r})$ we follow the procedure of von
Neumann \cite{reed, us1}, which requires the analysis of the square-integrable solutions of the equations
\begin{equation} \label{mles8}
  H^{\dagger}_r {\Psi}_{\pm} = \pm i {\Psi}_{\pm},
\end{equation}
where $\;  H^{\dagger}_r \; $ is the adjoint of $\;  H_r \; $
and $\;  {\Psi}_{\pm} \; $ are given by
\begin{equation} \label{mles9} 
 {\Psi}_{\pm} = \left( \begin{array}{c}
 {(u_{\pm} + v_{\pm}) ~ \frac{1}{\sqrt{2 \pi}} e^{i(j - \frac{1}{2})\phi} } \\
  {(u_{\pm} - v_{\pm}) ~ \frac{1}{\sqrt{2 \pi}} e^{i(j + \frac{1}{2})\phi} } \\
\end{array} \right) r^{\nu - \frac{1}{2}} e^{\mp r}
  \equiv 
   \left( \begin{array}{c} 
 {F_{\pm} ~ \frac{1}{\sqrt{2 \pi}} e^{i(j - \frac{1}{2})\phi} } \\
  {G_{\pm} ~ \frac{1}{\sqrt{2 \pi}} e^{i(j + \frac{1}{2})\phi} } \\
\end{array} \right).
\end{equation}
Note that $H_r$ and $H^{\dagger}_r$ have the same expressions as differential operators although their domains could be different. If the Eqs. (\ref{mles8}) have no square-integrable solutions, then the operator $\; H_{r} \; $ is self-adjoint in the domain $D(H_{r})$. However, if there are equal number(s) of square integrable solutions of Eqs. (\ref{mles8}), then the Dirac operator $\; H_{r} \; $ is not self-adjoint for the boundary conditions that define the domain $D(H_{r})$ in (\ref{mles7.2}). The latter could happen due to the effect of short range interactions induced by the impurity. If that happens, we would have to find a new set of boundary conditions, or equivalently a new domain for the radial Dirac operator in which it would be self-adjoint.  

We begin our analysis with Eqs. (\ref{mles8}), which lead to a set of coupled equations
\begin{equation} \label{mles10}
   r \frac{d u_{\pm}}{d r} + (\nu + i \beta \mp 2r) u_{\pm} - j v_{\pm} = 0,
\end{equation}
\begin{equation} \label{mles11}
  r \frac{d v_{\pm}}{d r} + (\nu - i\beta ) v_{\pm} - j u_{\pm} = 0
\end{equation}
for the functions $\; u_{\pm}, \; v_{\pm} \; $ appearing in
(\ref{mles9}).  

In terms of the variable $\; z = -2ikr, \; $ equations (\ref{mles10}) and
(\ref{mles11}) can be combined to give
\begin{equation} \label{mles12}
 z \frac{d^{2} v_{\pm} }{d z^{2}} + (1 + 2\nu - z)
 \frac{d v_{\pm}}{d z} - \left ( \nu - i \beta \right )
 v_{\pm} = 0. 
\end{equation}
Here it is implicitly understood that $ \; z = \pm 2r \;$ for $ \; k = \pm i, \;$
respectively.
As mentioned before, we have to search for square-integrable solutions of Eq.(\ref{mles8}).
We first look at the case with $ \; k = +i. \;$
The solution of (\ref{mles12}) which leads to a square integrable solution of
 (\ref{mles8}) is given by
\begin{equation} \label{mles13}
 v_+ = U(\nu - i \beta,~ 1+ 2\nu,~ z) = U(\nu - i \beta,~ 1+ 2\nu,~ 2r).
\end{equation}
With the help of Eq.(\ref{mles11}), this result for $\; v_{+ } \; $
further leads to
\begin{equation} \label{mles14}
 u_+ = \frac{\nu - i\beta}{j} \bigg [
    U(\nu - i \beta,~ 1+ 2\nu,~ 2r) - 2r U(\nu +1- i\beta,~ 2+ 2\nu,
   ~ 2r) \bigg ],
\end{equation}
so that we finally have
\begin{equation} \label{mles15}
 F_+ (r) = \bigg [ \frac{j+ \nu - i\beta}{j}
    U(\nu - i \beta,~ 1+ 2\nu,~ 2r) r^{\nu - \frac{1}{2}}
    - \frac{2(\nu - i \beta)}{j} U(\nu +1- i\beta,~ 2+ 2\nu,
   ~ 2r)  r^{\nu + \frac{1}{2}} \bigg ] e^{-r}
\end{equation}
for the radial part of the first component of the vector $\;
{\Psi}_{+} \; $ spanning the $\; k = +i \; $ deficiency subspace.
In deriving the expression for $\; u_+, \; $ we have used the relation
$\; U'(a,b,z) = -a U(a+1, b+1,z). \; $

 Next we want to find conditions
under which $\; F_+ \; $ and consequently $\; {\Psi}_{+} \; $
are square integrable. In the limit $\; r \rightarrow \infty, \;\; U(a,b,2r) \sim
r^{-a} \; $ and $\; F_+ \rightarrow 0, \; $ meaning that $\; F_+ \; $ is square
integrable at infinity. As $\; r \rightarrow 0, \; $ $\; M(a,b,2r) \rightarrow 1 \; $ 
and 
\begin{equation} \label{mles16}
  \int {|F_+|}^2 rdr \sim \int r^{-2 \nu} dr + \mbox{convergent terms}. 
\end{equation}
Therefore, for the range $\; 0 < \nu < \frac{1}{2}, \; $ $\; F_+ \; $ is a square integrable function. Similar analysis shows that for $\; 0 < \nu < \frac{1}{2}, \; $
 the entire radial wave-function is square integrable. Thus, for $\; k = +i \; $
we have a single square integrable solution of Eq.(\ref{mles8}). The deficiency index $n_+$ of $H_r$ is defined by the number of linearly independent and square integrable functions satisfying Eq.(\ref{mles8}) for $k = +i$. Our analysis shows that for gapless graphene with a charge impurity, $n_+ = 1$.

Let us now consider the case when $\; k = -i. \; $ In this situation
we have $\; z = -2ikr = -2r \; $ and a possible solution of
(\ref{mles12}) which leads to a square integrable solution of
 (\ref{mles8}) is given by
\begin{equation} \label{mles17}
 v_- = e^{z} U(1+ \nu + i \beta,~ 1+ 2\nu,~ -z) = e^{-2r} U(1+ \nu + i \beta,~ 1+ 2\nu,~ 2r).
\end{equation}
By using again Eq.(\ref{mles11}), this result for $\; v_{-} \; $
yields the following expression for $\; u_{-} \; $
\begin{equation} \label{mles18}
 u_- = \frac{\nu - i\beta}{j} e^{-2r} 
    U(1+ \nu + i \beta,~ 1+ 2\nu,~ 2r) -\frac{2r}{j}e^{-2r}
   \bigg [ U(1+ \nu + i\beta,~ 1+2 \nu,~ 2r) + (1+ \nu + i\beta ) 
     U(2+ \nu + i\beta,~ 2+ 2\nu,~ 2r)  \bigg ].
\end{equation}
If we make use of the recursive relation
 $\; U(a,b,z) -a U(a+1, b,z) - U(a, b-1,z) = 0, \; $
last expression for $\; u_{-} \; $ can be simplified to
\begin{equation} \label{mles19}
 u_- = \frac{\nu - i\beta}{j} e^{-2r} 
    U(1+ \nu + i \beta,~ 1+ 2\nu,~ 2r) -\frac{2r}{j}e^{-2r}
    U(1+ \nu + i\beta,~ 2+2 \nu,~ 2r),
\end{equation}
so that we finally have
\begin{equation} \label{mles20}
 F_- (r) =  e^{-r} \bigg [ \frac{j+ \nu - i\beta}{j}
     r^{\nu - \frac{1}{2}} U(1+ \nu + i \beta,~ 1+ 2\nu,~ 2r) 
    - \frac{2}{j}  r^{\nu + \frac{1}{2}} U(1+ \nu + i\beta,~ 2+ 2\nu,
   ~ 2r)  \bigg ]
\end{equation}
for the radial part of the first component of the vector $\;
{\Psi}_{-} \; $ spanning the $\; k = -i \; $ deficiency subspace.
By going through the similar procedure as before, we see that as
 $\; r \rightarrow \infty, \; F_- (r) \rightarrow 0. \; $ In the short
 distance limit we can again show that  $\; F_- \; $ and the
 corresponding entire radial wave-function is square integrable when 
$\; 0 < \nu < \frac{1}{2}. \; $ Thus, for $\; k = -i \; $ also, we
have a single square integrable solution to Eq.(\ref{mles8}) for the
range $\; 0 < \nu < \frac{1}{2} \; $ of the parameter $\; \nu, \; $
 implying that the deficiency index $\; n_- = 1\; $. 

According to von Neumann's analysis \cite{reed}, this result $\; n_+ = n_- =1, \; $ implies that $\; H_r \; $ itself is not self-adjoint on the domain $D(H_r),$  (\ref{mles7.2}), but can be made self-adjoint through a suitable choice of boundary conditions. When $n_+ = n_- = 1$, the allowed boundary conditions are labelled by a single real quantity, called the self-adjoint extension parameter. Our analysis thus shows that radial Dirac
operator $\; H_r \; $ admits a one parameter family of self-adjoint extensions when 
$\; 0 < \nu < \frac{1}{2}. \; $ When $\; n_+ = n_- =1, \; $ von Neumann's analysis  further says that domain in which $\; H_r \; $ is self-adjoint is given by $ \;{\mathcal{D}}_{z}(H_r) = {\mathcal{D}}(H_r)
\oplus  \{e^{i \frac{z}{2}} {\Psi}_{+} + e^{-i \frac{z}{2}} {\Psi}_{-}\}, \; $ where $ z \in R $ (mod $2 \pi$) is the self-adjoint extension parameter \cite{reed}. Physically, the domain ${\mathcal{D}}_{z}(H_r)$ provides the boundary conditions for which the radial Dirac operator for graphene is self-adjoint and we see that the boundary conditions are labelled by the parameter $z$.

In subsequent analysis it will be of interest to know the short
    distance behaviour of functions $\; F_+ \; $ and $\; F_-. \; $
 Their behaviour at short distances can be deduced with the help of the expression \cite{AS}
\begin{equation} \label{a5}
 U(a,b,z) = \frac{\pi}{\sin \pi b} \left [ \frac{M(a,b,z)}{\Gamma(1+a-b)\Gamma(b)} 
- z^{1-b} \frac{M(1+a-b,2-b,z)}{\Gamma(a)\Gamma(2-b)} \right ].
\end{equation}
Using (\ref{a5}), in the limit $r \rightarrow 0$, we have
\bea \label{mles21}
 F_+ &\longrightarrow& \frac{\pi}{\sin \pi (1+ 2\nu)}
   \frac{\nu + j -i\beta}{j} \frac{1}{\Gamma (-\nu - i\beta) \Gamma
   (1+ 2\nu)} r^{\nu - \frac{1}{2}} \\
&+& \bigg [ \frac{\pi}{\sin \pi (2+ 2\nu)} \frac{\nu - i\beta}{j}
   \frac{2^{-2\nu}}{\Gamma (1+\nu - i\beta) \Gamma (-2\nu)} -
   \frac{\pi}{\sin \pi (1+ 2\nu)} \frac{\nu +j - i\beta}{j}
   \frac{2^{-2\nu}}{\Gamma (\nu - i\beta) \Gamma (1-2\nu)} \bigg]
   r^{-\nu - \frac{1}{2}}, \nonumber
\eea
and 
\bea \label{mles22}
 F_- &\longrightarrow& \frac{\pi}{\sin \pi (1+ 2\nu)}
   \frac{\nu + j -i\beta}{j} \frac{1}{\Gamma (1-\nu + i\beta) \Gamma
   (1+ 2\nu)} r^{\nu - \frac{1}{2}} \\
&+& \bigg [ \frac{\pi}{\sin \pi (2+ 2\nu)} \frac{1}{j}
   \frac{2^{-2\nu}}{\Gamma (1+\nu + i\beta) \Gamma (-2\nu)} -
   \frac{\pi}{\sin \pi (1+ 2\nu)} \frac{\nu +j - i\beta}{j}
   \frac{2^{-2\nu}}{\Gamma (1+\nu + i\beta) \Gamma (1-2\nu)} \bigg] r^{-\nu - \frac{1}{2}}. \nonumber
\eea
This concludes our discussion of the allowed boundary conditions for the radial Dirac operator for the gapless graphene.

\subsection{Phase shifts and $S$ matrix for gapless graphene with a charge impurity}   

We now turn to problem of finding the scattering state solutions to Eq.(\ref{mles1}) for the parameter range $\; 0 < \nu < \frac{1}{2} \; $. This problem amounts to solving the set of coupled equations (\ref{mles6}) and (\ref{mles7}) or, equivalently, finding the solution of the 2nd order Eq.(\ref{mles7.1}).
The required solution of Eq.(\ref{mles7.1}) which leads to physical scattering states  has the form \cite{AS}
\begin{equation} \label{mles23}
 v = A_1 M \left ( \nu - i\beta,~ 1 + 2\nu,~z \right)
    + A_2 {z}^{- 2 \nu} M \left ( -\nu - i\beta,~ 1 -
    2\nu,~z \right),
\end{equation}
where $\; z= -2ikr. \;$

By assuming $v$ in the form (\ref{mles23}) and with the help of Eq.(\ref{mles7}), $u$ can be calculated as 
$$
  ju = A_1 z\frac{d}{dz}M \left ( \nu - i\beta,~ 1 + 2\nu,~z \right)
  + A_2 z(-2\nu)  z^{-2\nu -1}M \left ( -\nu - i\beta,~ 1 - 2\nu,~z \right)  
  + A_2  z^{-2\nu +1} \frac{d}{dz}M \left ( -\nu - i\beta,~ 1 - 2\nu,~z \right)
$$
\begin{equation} \label{mles24}
 + A_1 (\nu - i\beta) M \left ( \nu - i\beta,~ 1 + 2\nu,~z \right)
 + A_2 (\nu - i\beta) z^{-2\nu}M \left ( -\nu - i\beta,~ 1 - 2\nu,~z \right).
\end{equation}
After applying the recursive relation 
\begin{equation} \label{mles25}
  a M (a,b,z) + z M'(a,b,z) = a M (a+1,b,z),
\end{equation}
where prime denotes a derivation with respect to $z$, we get
\begin{equation} \label{mles26}
 ju = A_1 (\nu - i\beta )M \left (1+ \nu - i\beta,~ 1 + 2\nu,~z \right)
  + A_2  z^{-2\nu } (-\nu - i\beta ) M \left (1 -\nu - i\beta,~ 1 - 2\nu,~z \right).
\end{equation}
This leads to 
$$
 F(r) = r^{\nu - \frac{1}{2}}e^{ikr} 
  \bigg [ A_1 \frac{\nu - i\beta}{j} M \left (1 +\nu - i\beta,~ 1 + 2\nu,~z \right)
     + A_2 z^{-2\nu} \frac{-\nu - i\beta}{j} M \left (1 -\nu - i\beta,~ 1 - 2\nu,~z \right)
$$
\begin{equation} \label{mles27}
 + A_1 M \left ( \nu - i\beta,~ 1 + 2\nu,~z \right)
  + A_2  z^{-2\nu } M \left (-\nu - i\beta,~ 1 - 2\nu,~z \right) \bigg
  ],
\end{equation}
for the radial part of the first component of the wave-function
 $\; \Psi \; $ in (\ref{mles3}).

In order that the Hamiltonian provides a unitary evolution, the physical solution obtained above must belong to the domain of self-adjointness given by ${\mathcal{D}}_{z}(H_r)$. We will enforce this by matching the behaviour of the physical wavefunction with that of a typical element of ${\mathcal{D}}_{z}(H_r)$ in the limit $r \rightarrow 0$. In order to do this, note that as $\; r \rightarrow 0 \; $ 
\begin{equation} \label{mles28}
 F(r) \longrightarrow 
   A_1 \left ( 1 + \frac{\nu - i\beta}{j} \right ) r^{\nu -
   \frac{1}{2}}
   + A_2 {(-2ik)}^{-2\nu } \left ( 1 + \frac{-\nu - i\beta}{j} \right ) r^{-\nu -
   \frac{1}{2}}.
\end{equation}
In the same limit a  typical element of the domain ${\mathcal{D}}_{z}(H_r)$ is given by 
\begin{equation} \label{mles29}
   \Psi(r, \phi ) = \lambda \left
    ( e^{i \frac{z}{2}} {\Psi}_{+} + e^{-i \frac{z}{2}} {\Psi}_{-}
    \right ),
\end{equation}
where $\; \lambda \; $ is a constant. Comparing the upper component of the physical solution to that of an element of ${\mathcal{D}}_{z}(H_r)$ in the limit $\; r \rightarrow 0 \; $, we get 
\begin{equation} \label{mles30}
   A_1 = \lambda \frac{\pi}{\sin \pi (1+2\nu)} \left (
 \frac{ e^{i \frac{z}{2}}}{\Gamma (-\nu - i\beta ) \Gamma (1+2\nu)}
    + \frac{ e^{-i \frac{z}{2}}}{\Gamma (1-\nu + i\beta ) \Gamma (1+2\nu)}
    \right ),
\end{equation}
$$
   A_2 {(-2ik)}^{-2\nu} (j - \nu -i\beta) = - \lambda \frac{\pi}{\sin \pi (1+2\nu)} 
 \bigg [
   (\nu - i\beta) \frac{2^{-2\nu } e^{i \frac{z}{2}} }{\Gamma (1+\nu - i\beta ) \Gamma (-2\nu)} 
   + (\nu +j - i\beta) \frac{2^{-2\nu } e^{i \frac{z}{2}} }{\Gamma (\nu - i\beta ) \Gamma (1-2\nu)}  
$$ 
\begin{equation} \label{mles31}
 + \frac{2^{-2\nu } e^{-i \frac{z}{2}} }{\Gamma (1+\nu + i\beta ) \Gamma (-2\nu)}
  + (\nu +j - i\beta) \frac{2^{-2\nu } e^{-i \frac{z}{2}} }{\Gamma (1+\nu + i\beta ) \Gamma (1-2\nu)} 
 \bigg ].
\end{equation}

 We now proceed to find the phase shifts characterizing the scattering
 process in gapless graphene. These quantities are of interest to us because it
 is possible to extract important informations out of them, which in
 turn are related to various physical properties of graphene such as
 the electrical conductivity and transport properties.
    In order to find the phase shifts and the scattering matrix, we
 have to investigate the asymptotic behaviour of the wave-function
 $\; \Psi, \; $ Eq.(\ref{mles3}), or equivalently, of the function $\; F(r) \; $
 given in the relation (\ref{mles27}). To do this,
 we note that 
the behaviour of the confluent hypergeometric function $M (a,b,z)$ in the asymptotic region ${\rm Re} (z)=0$ and ${\rm
Im} (z) \rightarrow +\infty $ is given by \cite{AS}
\begin{equation} \label{9}
  M (a,b,z) \longrightarrow \frac{\Gamma
 (b)}{\Gamma (a)} e^z z^{a-b} \bigg [ 1+ O({|z|}^{-1}) \bigg ] +
 \frac{\Gamma (b)}{\Gamma (b - a)} {z}^{-a} e^{\pm i \pi a} \bigg [ 1+ O({|z|}^{-1})
 \bigg ] ,
\end{equation}
Due to the fact that
we are dealing with the problem where $\; {\rm Re} (z) = 0, \;$ both
leading terms in the asymptotic expansion (\ref{9}) of $M$
approximately have the contribution of the same order, so that both of
them have to be taken into account.
Knowing that, the  behaviour of $\; F(r) \; $ when $ \; r \rightarrow
\infty \; $ is now given by

\bea \label{mles33}
  F(r) &{\longrightarrow}& 
   {(-2ik)}^{-\nu } {(-i)}^{-i\beta} \left (
  A_1  \frac{\nu - i\beta}{j} 
      \frac{\Gamma (1+2\nu)}{\Gamma (1+\nu - i\beta )} 
   + A_2 \frac{-\nu - i\beta}{j}  \frac{\Gamma (1-2\nu)}{\Gamma (1-\nu - i\beta )} 
     \right ) \frac{e^{-i (kr + \beta \ln 2kr)}}{\sqrt{r}} \nonumber \\
&+&  {(-2ik)}^{-\nu } {(-i)}^{ i\beta} \left ( A_1 
      \frac{\Gamma (1+2\nu)}{\Gamma (1+\nu + i\beta )} 
    e^{-i \pi (\nu - i\beta)}
   + A_2  \frac{\Gamma (1-2\nu)}{\Gamma (1-\nu + i\beta )} 
     e^{i \pi (\nu + i\beta)}  \right )
  \frac{e^{i (kr + \beta \ln 2kr)}}{\sqrt{r}}.
\eea
The scattering matrix and the corresponding phase shift can be written as 
\begin{equation} \label{mles34}
   S(k) \equiv e^{2i \delta (k)}     e^{-4\pi \beta}
   \frac{\frac{A_1}{A_2} \frac{\Gamma (1+2\nu)}{\Gamma (1+\nu +
   i\beta )} e^{-i \pi \nu } +  \frac{\Gamma (1-2\nu)}{\Gamma (1-\nu +
   i\beta )} e^{i \pi \nu }}{\frac{A_1}{A_2}
    \frac{\nu - i\beta}{j} \frac{\Gamma (1+2\nu)}{\Gamma (1+\nu -
   i\beta )} + \frac{-\nu - i\beta}{j} \frac{\Gamma (1-2\nu)}{\Gamma (1-\nu -
   i\beta )} },
\end{equation}
where the ratio $\; A_1 / A_2 \; $ can be obtained from Eqs.(\ref{mles30}) and (\ref{mles31})
and is given by
\begin{equation} \label{mles35}
 \frac{A_1}{A_2} = 
   - {(-ik)}^{-2\nu}
    \frac{\Gamma (1-2\nu)}{\Gamma (1+2\nu)}
   \frac{ \frac{ e^{i \frac{z}{2}}}{\Gamma (-\nu - i\beta )}
    + \frac{ e^{-i \frac{z}{2}}}{\Gamma (1-\nu + i\beta ) }}
    { \frac{ e^{i \frac{z}{2}} }{\Gamma (\nu - i\beta )} + 
   \frac{ e^{-i \frac{z}{2}} }{\Gamma (1+\nu + i\beta ) }}.
\end{equation}

It should be emphasized that the above analysis is relevant only when the system parameters are such that $0 < \nu < \frac{1}{2}$, where $\nu = \sqrt{j^2 - \beta^2}$. For this range of the system parameters, the radial Dirac operator $H_r$ in Eq. (\ref{raddirac}) is not self-adjoint with the boundary conditions describing the domain $D(H_{r})$ (\ref{mles7.2}). The allowed set of more general boundary conditions for which the radial Dirac operator $H_r$ is self-adjoint is labelled by a real parameter $z$ defined modulo $2 \pi$.  The above analysis clearly shows that the spectral data including the phase shifts and the scattering matrix explicitly depend on the boundary condition through the self-adjoint extension parameter $z$. For each value of $z$, we have an inequivalent quantum description of the gapless graphene system. This is not surprising as in our framework, the boundary conditions (or equivalently the self-adjoint extension parameter $z$ ) encode the effects of the short range interactions due to the impurity. Different values of $z$ correspond to different combined effects of the short range interactions, leading to inequivalent physical results. It should be noted that the value of the self-adjoint parameter $z$ cannot be fixed by the above analysis alone. For any given system of gapless graphene with a charge impurity, the value of the parameter $z$ would have to be fixed empirically by measuring physical quantities which depend on the scattering data, such as resistivity \cite{kats}. We shall make further comments about this in the conclusions.

We end this Section by noting that the $S$ matrix in (\ref{mles34}) does not have any poles in the positive imaginary value of the energy. This implies that there are no bound states for a gapless graphene in the presence of a subcritical strength charge impurity, in agreement with the Klein paradox.


\section{Scattering in gapped graphene with a charge impurity}

In this Section we shall discuss the scattering states, phase shifts and the $S$ matrix for a gapped graphene in the presence of a charge impurity. We start with a brief review of the gapped graphene system where a charge impurity is present. This is followed by a review of the allowed boundary conditions for which the gapped graphene Hamiltonian admits a unitary evolution \cite{us2}. Finally we find the scattering data for the gapped graphene system which under certain conditions depend on an additional parameter that labels the boundary conditions. We also comment on how to determine this parameter empirically. 

\subsection{Gapped graphene with a charge impurity}

In this case, the 2D Dirac equation can be written as 
\begin{equation} \label{1}
  H \Psi = E \Psi,
\end{equation}
 where the Hamiltonian is given by
\begin{equation} \label{2}
  H = -i ( {\sigma}_{1} {\partial}_{x} + {\sigma}_{2} {\partial}_{y} )
    + m {\sigma}_{3} - \frac{\alpha}{r}
\end{equation}
and the eigenfunctions are two-component wavefunctions of the form
\begin{equation} \label{separ} 
 \Psi (r, \phi) = \left( \begin{array}{c}
 {\psi_1 (r ) ~ \frac{1}{\sqrt{2 \pi}} e^{i(j - \frac{1}{2})\phi} } \\
  {i \psi_2 (r )~ \frac{1}{\sqrt{2 \pi}} e^{i(j + \frac{1}{2})\phi} } \\
\end{array} \right).
\end{equation}
In the above equations $\; r \; $ and $\; \phi \; $ are the radial
variable and the angle in the $\; x-y \; $ plane, respectively.
Consider the ansatz
\begin{equation} \label{3}
 {\psi}_{1}(r) = 
   \sqrt{m + E} ~ e^{-\frac{\rho}{2}} {\rho}^{\nu - \frac{1}{2}}
    (P(\rho) + Q(\rho) ), 
\end{equation}
\begin{equation} \label{4}
 {\psi}_{2}(r) = 
   \sqrt{m - E} ~ e^{-\frac{\rho}{2}} {\rho}^{\nu - \frac{1}{2}}
    (P(\rho) - Q(\rho) ), 
\end{equation}
 where $\; \rho = 2 \gamma r = 2 \sqrt{{m}^2 - {E}^2} ~ r \; $ and $\; \nu = \sqrt{{j}^2 -
 {\alpha}^2}, \;\;\; j \; $ being a half integer. After inserting the ansatz
 (\ref{3}), (\ref{4}) into (\ref{1}), we obtain the set of coupled equations 
\begin{equation} \label{rad}
H_\rho \left( \begin{array}{c}
P  \\
Q \\
\end{array} \right) = \left( \begin{array}{cc}
\rho \frac{d}{d \rho} + \nu - \frac{\alpha E}{\gamma} & - j + \frac{m \alpha}{\gamma}  \\
- j - \frac{m \alpha}{\gamma} & \rho \frac{d}{d \rho} + \nu - \rho + \frac{\alpha E}{\gamma} \\
\end{array} \right)
\left( \begin{array}{c}
P  \\
Q  \\
\end{array} \right) = 0,
\end{equation}
where $H_\rho$ defined above denotes the radial Dirac operator. These equations can be decoupled to give
\begin{equation} \label{5}
 \rho \frac{d^{2} P}{d {\rho}^{2}} + (1 + 2\nu - \rho)
 \frac{dP}{d \rho} - \left ( \nu - \frac{\alpha E}{\gamma} \right )
 P = 0, 
\end{equation}
\begin{equation} \label{6}
 \rho \frac{d^{2} Q}{d {\rho}^{2}} + (1 + 2\nu - \rho)
 \frac{d Q}{d \rho} - \left (1 + \nu - \frac{\alpha E}{\gamma} \right )
 Q = 0.  
\end{equation}
These equations can be solved in terms of confluent hypergeometric functions \cite{AS}. We now proceed to discuss the allowed boundary conditions for this problem.


\subsection{Boundary conditions for gapped graphene with a charge impurity}

The analysis for the allowed boundary conditions for gapped graphene with a charge impurity was already discussed in our previous paper \cite{us2}. Here we review that work for completeness and also for obtaining the formulae that would be useful in the subsequent analysis. In order to find the deficiency indices $n_{\pm}$ for $H_\rho$ and
 subsequently for $H,$
 we need to solve equations
\begin{equation} \label{a1} 
  H^{\dagger} {\Psi}_{\pm} = \pm i
    {\Psi}_{\pm},
\end{equation}
where $H^{\dagger}$ has the same differential expression as for $H$ in Eq. (\ref{2}), although their domains could be different.
In Eq.(\ref{a1}) $\; {\Psi}_{\pm}  \;$ are two-component spinors of the form
\begin{equation} \label{a2} 
 {\Psi}_{\pm} = \left( \begin{array}{c}
 {{\psi}_{1 \pm} (r ) ~ \frac{1}{\sqrt{2 \pi}} e^{i(j - \frac{1}{2})\phi} } \\
  {i {\psi}_{2 \pm} (r )~ \frac{1}{\sqrt{2 \pi}} e^{i(j + \frac{1}{2})\phi} } \\
\end{array} \right),
\end{equation}
where
\begin{eqnarray} \label{a2.0}
\psi_{1\pm}  &=& \sqrt{m \pm i} e^{-\frac{\rho}{2}} \rho^{\nu -
  \frac{1}{2}} (P_\pm + Q_\pm ) (\rho), \\
\psi_{2\pm}  &=& \sqrt{m \mp i} e^{-\frac{\rho}{2}} \rho^{\nu -
  \frac{1}{2}} (P_\pm - Q_\pm ) (\rho).
\end{eqnarray}
The functions $\psi_{1\pm}$ and $ \psi_{2\pm}$ have to be square
 integrable in $R^+$ with a measure $\rho d \rho$.
To get any further information  one has to solve
 for $P_{\pm}$ and $Q_{\pm}$ from the equations
\begin{equation} \label{a2.1}
   \rho \frac{d P_{\pm}}{d \rho} + (\nu - \frac{i \alpha }{{\gamma}_{\pm}}) P_{\pm}
  + \left ( \frac{m \alpha }{{\gamma}_{\pm}} - j \right ) Q_{\pm} = 0,
\end{equation}
\begin{equation} \label{a2.2}
  \rho \frac{d Q_{\pm}}{d \rho} + (\nu - \rho + \frac{i \alpha }{{\gamma}_{\pm}}) Q_{\pm}
  - \left ( j + \frac{m \alpha }{{\gamma}_{\pm}}  \right ) P_{\pm} = 0,
\end{equation}
 where $ \; \gamma_{\pm} = \sqrt{m^2 + 1} \;$ and $ \; \rho = 2
 {\gamma}_{\pm} r \;$.
These equations can also be decoupled to give
\begin{eqnarray} \label{a3}
\rho \frac{d^2 P_{\pm}}{d \rho^2} + (1 + 2 \nu -\rho)\frac{d P_{\pm}}{d \rho}
 - \left ( \nu - \frac{i \alpha }{\gamma_{\pm}} \right )P_{\pm} &=& 0 \\
\rho \frac{d^2 Q_{\pm}}{d \rho^2} + (1 + 2 \nu -\rho)
\frac{d Q_{\pm}}{d \rho} - \left (1 + \nu - \frac{i \alpha }{\gamma_{\pm}} \right )Q_{\pm} &=& 0,
\end{eqnarray}
Let us first look at the deficiency subspace determined by the positive
 sign in the above expression. Then
 the required solution of Eq.(\ref{a3}) is
\begin{equation} \label{a4}
P_+ = U \left ( \nu - \frac{i \alpha}{\gamma_+}, 1 + 2 \nu, \rho \right ), 
\end{equation}
where $U$ denotes a confluent hypergeometric function defined in Eq. (\ref{a5}).
 Since $Q_+ $ is not independent of $P_+, $ it can be calculated with
 the help of the relation (\ref{a2.1}),
\begin{equation} \label{a5.1}
 \left ( j - \frac{m \alpha}{{\gamma}_+} \right ) Q_+  =  \rho \frac{d}{d \rho} 
 U \left ( \nu - \frac{i \alpha}{\gamma_+}, 1 + 2 \nu, \rho \right )
  +  \left ( \nu - \frac{i \alpha}{\gamma_+} \right )
    U \left ( \nu - \frac{i \alpha}{\gamma_+}, 1 + 2 \nu, \rho \right ). 
\end{equation}
By using the differential recursive relation
   $\; z U'(a,b,z) + a U(a,b,z) = a(1+a-b) U(a + 1,b,z),  \; $
with prime denoting a derivation with respect to $z,$ we get
\begin{equation} \label{a5.2}
   \left ( j - \frac{m \alpha}{{\gamma}_+} \right ) Q_+ = \left ( \nu - \frac{i \alpha}{{\gamma}_+} \right ) 
    \left (- \nu - \frac{i \alpha}{{\gamma}_+} \right ) 
    U \left (1+ \nu - \frac{i \alpha}{\gamma_+}, 1 + 2 \nu, \rho \right ). 
\end{equation}
In the limit $\rho \longrightarrow 0$ 
the functions (\ref{a4}) behave as
\begin{eqnarray} \label{a6}
P_+ &\longrightarrow& a (A_+ - B_+ \rho^{-2 \nu}), \\
Q_+ &\longrightarrow& a (C_+ - D_+ \rho^{-2 \nu}),
\end{eqnarray}
where $a = \frac{\pi}{\sin \pi (1+ 2 \nu)} $ and
\begin{eqnarray} \label{a7}
A_+ &=& \frac{1}{\Gamma(-\nu - \frac{i \alpha}{\gamma_+}) \Gamma(1 + 2 \nu)} ~~~~~~~
B_+ = \frac{1}{\Gamma(\nu - \frac{i \alpha}{\gamma_+}) \Gamma(1 - 2 \nu)} \\
C_+ &=& \frac{( \nu - \frac{i \alpha}{{\gamma}_+} )
 (- \nu - \frac{i \alpha}{{\gamma}_+})}{( j - \frac{m \alpha}{{\gamma}_+} )}
   \frac{1}{\Gamma(1 -\nu - \frac{i \alpha}{\gamma_+}) \Gamma(1 + 2 \nu)} ~~~~
D_+ = \frac{( \nu - \frac{i \alpha}{{\gamma}_+} )
 (- \nu - \frac{i \alpha}{{\gamma}_+})}{( j - \frac{m \alpha}{{\gamma}_+} )}
 \frac{1}{\Gamma(1 + \nu - \frac{i \alpha}{\gamma_+}) \Gamma(1 - 2 \nu)} 
\end{eqnarray}
are constants depending on the system parameters. We can now easily show that $\psi_{1+}$ and $\psi_{2+}$ are square integrable everywhere provided $0 < \nu < \frac{1}{2}$ \cite{us2}. Thus $n_+=1$ for the parameter range $0 < \nu < \frac{1}{2}$.

   In a similar way, by analyzing the deficiency subspace
   characterized by the negative sign in (\ref{a1}), we obtain
\begin{equation} \label{a7.1}
 P_- = U \left ( \nu + \frac{i \alpha}{\gamma_+}, 1 + 2 \nu, \rho \right ), 
\end{equation}
\begin{equation} \label{a7.2}
   \left ( j - \frac{m \alpha}{{\gamma}_+} \right ) Q_- = \left ( \nu + \frac{i \alpha}{{\gamma}_+} \right ) 
    \left (- \nu + \frac{i \alpha}{{\gamma}_+} \right ) 
    U \left (1+ \nu + \frac{i \alpha}{\gamma_+}, 1 + 2 \nu, \rho \right ). 
\end{equation}
In addition,
in the limit $\rho \longrightarrow 0$ 
the functions $P_-$ and $Q_-$ behave as
\begin{eqnarray} \label{A7.3}
P_- &\longrightarrow& a (A_- - B_- \rho^{-2 \nu}), \\
Q_- &\longrightarrow& a (C_- - D_- \rho^{-2 \nu}),
\end{eqnarray}
where $a = \frac{\pi}{\sin \pi (1+ 2 \nu)} $ and
\begin{eqnarray} \label{a7.4}
A_- &=& \frac{1}{\Gamma(-\nu + \frac{i \alpha}{\gamma_+}) \Gamma(1 + 2 \nu)} ~~~~~~~
B_- = \frac{1}{\Gamma(\nu + \frac{i \alpha}{\gamma_+}) \Gamma(1 - 2 \nu)} \\
C_- &=& \frac{( \nu + \frac{i \alpha}{{\gamma}_+} )
 (- \nu + \frac{i \alpha}{{\gamma}_+})}{( j - \frac{m \alpha}{{\gamma}_+} )}
   \frac{1}{\Gamma(1 -\nu + \frac{i \alpha}{\gamma_+}) \Gamma(1 + 2 \nu)} ~~~~
D_- = \frac{( \nu + \frac{i \alpha}{{\gamma}_+} )
 (- \nu + \frac{i \alpha}{{\gamma}_+})}{( j - \frac{m \alpha}{{\gamma}_+} )}
 \frac{1}{\Gamma(1 + \nu + \frac{i \alpha}{\gamma_+}) \Gamma(1 - 2 \nu)}. 
\end{eqnarray}
From this we see that relations 
\begin{equation} \label{a7.5}
 A_- = {\bar{A}}_+, \;\;\;\; B_- = {\bar{B}}_+, \;\;\;\;
    C_- = {\bar{C}}_+, \;\;\;\;  D_- = {\bar{D}}_+ \;\;\;\; 
\end{equation}
hold. Similar analysis as before shows that $n_-=1$ for the parameter range $0 < \nu < \frac{1}{2}$ as well. Thus for the gapped graphene with a charge impurity, $n_+ = n_- =1$ when $0 < \nu < \frac{1}{2}$. Therefore, this system admits a one parameter family of self-adjoint extensions for $0 < \nu < \frac{1}{2}$ \cite{us2}.


\subsection{Phase shifts and $S$ matrix for gapped graphene with a charge impurity}

Scattering states correspond to solutions of Eq.(\ref{1}) when $ E > m.$
In this case the parameter $ \; \gamma = \sqrt{m^{2} - E^{2} } \; $ becomes purely
imaginary, i.e.  $ \; \gamma = iq, \;$ where the real parameter $q$ is defined
as $q= \sqrt{E^{2} - m^{2}} $. Consequently, the variable $ \; \rho
\;$ also becomes purely imaginary, $ \; \rho = 2 iqr.$

In order to analyze the $r \rightarrow
\infty$ limit of the scattering states, we first express the solutions
written in terms of the hypergeometric function $U$ in terms of the
hypergeometric function $M$ using Eq. (\ref{a5}) and then carry out
the asymptotic expansion (\ref{9}).
 As was the case for the gapless graphene, here again both
leading terms in the asymptotic expansion (\ref{9}) of $M$
approximately have the contribution of the same order, so that both of
them have to be taken into account.

The solutions of Eqs.(\ref{5}), (\ref{6}) which lead to physical scattering states 
 are found to have the following form
\begin{equation} \label{11}
 P(\rho) = A M \left ( \nu - \frac{\alpha E}{\gamma}, 1 + 2\nu, \rho \right)
    + B {\rho}^{- 2 \nu} M \left ( -\nu - \frac{\alpha E}{\gamma}, 1 -
    2\nu, \rho \right),
\end{equation}
\begin{equation} \label{12}
 Q(\rho) = C M \left (1 + \nu - \frac{\alpha E}{\gamma}, 1 + 2\nu,
    \rho \right )
    + D {\rho}^{- 2 \nu} M \left (1 -\nu - \frac{\alpha E}{\gamma}, 1
    - 2\nu, \rho \right ) .
\end{equation}
 In this case the solution for the function $ {\psi}_{1}, $ appearing in
(\ref{separ}) would look like
$$
 {\psi}_{1}(r) = \sqrt{m + E} ~ e^{-\frac{\rho}{2}} {\rho}^{\nu - \frac{1}{2}}
 \bigg [ A(q)M \left ( \nu - \frac{\alpha E}{\gamma}, 1 + 2\nu, \rho
  \right )
    + B(q) {\rho}^{- 2 \nu} M \left ( -\nu - \frac{\alpha E}{\gamma},
 1 - 2\nu, \rho \right )
$$
\begin{equation} \label{13}
 + C(q) M \left (1 + \nu - \frac{\alpha E}{\gamma}, 1 + 2\nu, \rho \right )
    + D(q) {\rho}^{- 2 \nu} M \left (1 -\nu - \frac{\alpha E}{\gamma},
 1 - 2\nu, \rho \right ) \bigg ],
\end{equation}
where we have assumed that the coefficients
$A(q)$ and $B(q)$ depend on the real parameter $q$.
However, at this stage, it is important to stress
 that functions $P$ and $Q$ are not independent of each other,
but rather they are
   related through the set of coupled equations
\begin{equation} \label{13.1}
   \rho \frac{d P}{d \rho} + (\nu - \frac{\alpha E}{\gamma}) P
  + \left ( \frac{m \alpha }{\gamma} - j \right ) Q = 0,
\end{equation}
\begin{equation} \label{13.2}
  \rho \frac{d Q}{d \rho} + (\nu - \rho + \frac{\alpha E}{\gamma}) Q
  - \left ( j + \frac{m \alpha }{\gamma}  \right ) P = 0.
\end{equation}
As a consequence, the
constants $A(q), \; B(q), C(q) \; $ and $\; D(q) \; $  are also not all independent.

    By assuming that $P$ has the form (\ref{11}), we can obtain $Q$ as 
\bea
 &&\left ( j - \frac{m \alpha }{\gamma}  \right ) Q = 
\left ( \nu - \frac{\alpha E}{\gamma} \right)
    \bigg [ 
   A M \left ( \nu - \frac{\alpha E}{\gamma}, 1 + 2\nu, \rho \right) 
   +  B {\rho}^{- 2 \nu } M \left (- \nu - \frac{\alpha E}{\gamma}, 1 - 2\nu, \rho \right) 
   \bigg ] +  \\
&& \rho \bigg [ 
   A M' \left ( \nu - \frac{\alpha E}{\gamma}, 1 + 2\nu, \rho \right) 
   +  B {\rho}^{- 2 \nu } M' \left (- \nu - \frac{\alpha E}{\gamma}, 1 - 2\nu, \rho \right) 
   + B (- 2 \nu ) {\rho}^{- 2 \nu - 1 }
 M \left (- \nu - \frac{\alpha E}{\gamma}, 1 - 2\nu, \rho \right)
   \bigg ] \nonumber
\eea 
where prime denotes the derivation with respect to $ \rho. $
By applying the recursive relation 
\begin{equation} \label{13.4}
  a M (a,b,z) + z M'(a,b,z) = a M (a+1,b,z)
\end{equation}
(prime denotes derivation with respect to $z$)
we obtain
\begin{equation} \label{13.7}
  \left ( j - \frac{m \alpha }{\gamma}  \right ) Q = A(q) \left ( \nu - \frac{\alpha E}{\gamma} \right)
     M \left (1 + \nu - \frac{\alpha E}{\gamma}, 1 + 2\nu, \rho \right)
  + B(q) {\rho}^{- 2 \nu }  
    \left (- \nu - \frac{\alpha E}{\gamma} \right)
    M \left (1 - \nu - \frac{\alpha E}{\gamma}, 1 - 2\nu, \rho \right). 
\end{equation}
   Since $Q$ from (\ref{13.7}) has to  coincide with $ Q$ from (\ref{12}),
  (at least up to a constant), we must have 
\begin{equation} \label{13.9}
  C(q) = A(q) \frac{\nu - \frac{\alpha E}{\gamma}}{j - \frac{m \alpha
 }{\gamma} },
\end{equation} 
\begin{equation} \label{13.10}
 D(q) = B(q) \frac{-\nu - \frac{\alpha E}{\gamma}}{j - \frac{m \alpha
 }{\gamma} }.
\end{equation} 

 As $ r \rightarrow \infty $, using (\ref{9}) and noting that $\gamma = i q$, the $ {\psi}_{1} $ component of the wave function has the form
\bea \label{17}
 {\psi}_{1}(r)  &\longrightarrow& 
   \sqrt{m + E} ~ {(2iq)}^{ -\frac{1}{2} -i \frac{\alpha E}{q}}
       \bigg [ A(q) \frac{\Gamma (1 + 2\nu)}{\Gamma (1 + \nu -i \frac{\alpha
   E}{q})}  e^{-i \pi (\nu + i \frac{\alpha E}{q})} +
  B(q) \frac{\Gamma (1 - 2\nu)}{\Gamma (1- \nu -i \frac{\alpha  E}{q})} 
   e^{-i \pi (- \nu +i \frac{\alpha E}{q} )} \bigg ] 
 \frac{e^{-i ( qr + \frac{\alpha E}{q} \ln r )}}{\sqrt{r}} \nonumber \\
&+& \sqrt{m + E} ~ {(2iq)}^{- \frac{1}{2} +i \frac{\alpha E}{q}}
    \bigg [ C(q) \frac{\Gamma (1 + 2\nu)}{\Gamma (1 + \nu + i \frac{\alpha  E}{q})}    
  + D(q) \frac{\Gamma (1 - 2\nu)}{\Gamma (1- \nu + i \frac{\alpha    E}{q})} 
   \bigg ]  \frac{e^{i ( qr + \frac{\alpha E}{q} \ln r )}}{\sqrt{r}},
\eea
from which one can identify incoming and outgoing waves and the
corresponding amplitudes.
The scattering matrix and the phase shift can be obtained
 from the limiting form (\ref{17}) of the wave function as a ratio of
its outgoing and incoming amplitude,
\begin{equation} \label{18}
  S (q) =  e^{2i \delta (q)} 
   = {(2iq)}^{2 i \frac{\alpha E}{q}}
 \frac{ C(q) \frac{\Gamma (1 + 2\nu)}{\Gamma (1 + \nu + i \frac{\alpha  E}{q})}    
  + D(q) \frac{\Gamma (1 - 2\nu)}{\Gamma (1- \nu + i \frac{\alpha
   E}{q})}} { A(q) \frac{\Gamma (1 + 2\nu)}{\Gamma (1 + \nu -i \frac{\alpha
   E}{q})} e^{-i \pi ( \nu +i \frac{\alpha E}{q} )}  +
  B(q) \frac{\Gamma (1 - 2\nu)}{\Gamma (1- \nu -i \frac{\alpha  E}{q})} e^{-i \pi (- \nu +i \frac{\alpha E}{q} )} }.
\end{equation}
So far the constants
$A(q),$ $B(q),$ $C(q)$ and $ D(q) $ are not completely specified.
All we know about them is that they  are related through the relations (\ref{13.9}),(\ref{13.10})
  and that they depend on the transfer momentum $ q.$
To find a further relationship between them, it is of interest to look 
at the short distance limit, $\; r \rightarrow 0, \; $ of the function (\ref{13})
\begin{equation} \label{19}
  {\psi}_{1}(r)  \longrightarrow 
  \sqrt{m + E} \left ( (A(q) + C(q)) {\rho}^{\nu - \frac{1}{2}}
          + (B(q) + D(q)) {\rho}^{-\nu - \frac{1}{2}} \right ).
\end{equation}
 This expression gives us the function $ {\psi}_{1} $ to lowest order in $r.$

We already know that the radial Dirac operator $H_{\rho}$
and consequently Hamiltonian $H,$ (\ref{2}), admits a one parameter family of self-adjoint extension when $ \; 0 < \nu < \frac{1}{2}. \; $  In this case, the domain of self-adjointness of the Hamiltonian (\ref{2}) is given by 
 ${\mathcal{D}}_{z}(H) = {\mathcal{D}}(H)
\oplus  \{e^{i \frac{z}{2}} {\Psi}_{+} + e^{-i \frac{z}{2}} {\Psi}_{-}
\}.$ In the limit $\; r \rightarrow 0 ,\; $  a typical element of the domain ${\mathcal{D}}_{z}(H)$ can be written as
\begin{equation} \label{20}
   \Psi(r, \phi ) = c \left
    ( e^{i \frac{z}{2}} {\Psi}_{+} + e^{-i \frac{z}{2}} {\Psi}_{-}
    \right ),
\end{equation}
where $c$ is some constant and ${\Psi}_{\pm}$ are square integrable
solutions of Eqs. (\ref{a1}).

After we make use of the relation (\ref{20}) and the fact
 that the coefficients of appropriate
powers of $r$ at both sides in (\ref{20})  must match, we get the following two
conditions 
\begin{equation} \label{21}
    {(2 \gamma )}^{\nu - \frac{1}{2}}
   \left ( A(q) + C(q) \right ) \sqrt{m + E} =  ca \left ( \sqrt{m + i} e^{i \frac{z}{2}}
    (A_{+} + C_{+}) + \sqrt{m - i} e^{-i \frac{z}{2}}
    ({\bar{A}}_{+} + {\bar{C}}_{+}) \right ) {(2 {\gamma}_{+} )}^{\nu - \frac{1}{2}}
\end{equation}
\begin{equation} \label{22}
  {(2 \gamma )}^{-\nu - \frac{1}{2}}
   \left ( B(q) + D(q) \right ) \sqrt{m + E} =  -ca \left ( \sqrt{m + i} e^{i \frac{z}{2}}
    (B_{+} + D_{+}) + \sqrt{m - i} e^{-i \frac{z}{2}}
    ({\bar{B}}_{+} + {\bar{D}}_{+}) \right ) {(2 {\gamma}_{+} )}^{-\nu - \frac{1}{2}}, 
\end{equation}
where $\; z \; $ is the self-adjoint extension parameter
and  all other quantities appearing in the above relations are defined in the previous section.
The last two equations yield
\begin{equation} \label{23}
   \frac{A(q) + C(q)}{B(q) + D(q)}  =  - \frac{\sqrt{m + i} e^{i \frac{z}{2}}
    (A_{+} + C_{+}) + \sqrt{m - i} e^{-i \frac{z}{2}}
    ({\bar{A}}_{+} + {\bar{C}}_{+})}
  {\sqrt{m + i} e^{i \frac{z}{2}}
    (B_{+} + D_{+}) + \sqrt{m - i} e^{-i \frac{z}{2}}
    ({\bar{B}}_{+} + {\bar{D}}_{+})}
   {(2 {\gamma}_{+} )}^{2\nu } {(2 \gamma )}^{-2 \nu }
= - \frac{\xi_1 \cos(\theta_1 + \frac{z}{2})}
  {\xi_2 \cos(\theta_2 + \frac{z}{2})}
   {(2 {\gamma}_{+} )}^{2\nu } {(2 \gamma )}^{-2 \nu },
\end{equation}
where we have defined  $ \; \sqrt{m + i}(A_+ + C_+) \equiv \xi_1 e^{i \theta_1} \; $
 and $\; \sqrt{m + i}(B_+ + D_+) \equiv \xi_2 e^{i \theta_2}. \; $ Using (\ref{13.9}), (\ref{13.10}) and (\ref{23}), the $S$ matrix in (\ref{18}) can be written as 
\begin{equation} \label{25}
  S (q) =  
    {(2iq)}^{2 i \frac{\alpha E}{q}}
 \frac{ - \frac{\xi_1 \cos(\theta_1 + \frac{z}{2})}
  {\xi_2 \cos(\theta_2 + \frac{z}{2})}
   {(2 {\gamma}_{+} )}^{2\nu } {(2 \gamma )}^{-2 \nu }
  \frac{ 1 + {\omega}_{2}}{1 + {\omega}_{1}} {\omega}_{1}
 \frac{\Gamma (1 + 2\nu)}{\Gamma (1 + \nu + i \frac{\alpha  E}{q})}    
  + {\omega}_{2} \frac{\Gamma (1 - 2\nu)}{\Gamma (1- \nu + i \frac{\alpha
   E}{q})}}
 { - \frac{\xi_1 \cos(\theta_1 + \frac{z}{2})}
  {\xi_2 \cos(\theta_2 + \frac{z}{2})}
   {(2 {\gamma}_{+} )}^{2\nu } {(2 \gamma )}^{-2 \nu }
  \frac{ 1 + {\omega}_{2}}{1 + {\omega}_{1}}
 \frac{\Gamma (1 + 2\nu)}{\Gamma (1 + \nu -i \frac{\alpha
   E}{q})} e^{-i \pi ( \nu +i \frac{\alpha E}{q} )}  +
   \frac{\Gamma (1 - 2\nu)}{\Gamma (1- \nu -i \frac{\alpha  E}{q})} e^{-i \pi ( - \nu +i \frac{\alpha E}{q} )}
   },
\end{equation}
where
\begin{equation} \label{28}
 {\omega}_1 \equiv \frac{C(q)}{A(q)} = \frac{\nu - \frac{\alpha E}{\gamma}}{j - \frac{m \alpha }{\gamma} },
 \quad  {\omega}_2 \equiv \frac{D(q)}{B(q)} =  \frac{-\nu - \frac{\alpha E}{\gamma}}{j - \frac{m \alpha }{\gamma} }.    
\end{equation}

The expression in (\ref{25}) gives the $S$ matrix for gapped graphene
for the parameter range $\; 0 < \nu < \frac{1}{2} \; $. For this range
of $\nu$, the appropriate boundary conditions for which the
Hamiltonian (\ref{2}) is self-adjoint and the corresponding time evolution unitary requires the introduction of an additional real self-adjoint extension parameter $z$, which labels the allowed boundary conditions. The phase shifts and the $S$ matrix depend explicitly on $z$. For each value of $z$ (mod $2 \pi$), we have an inequivalent set of the scattering data. The above analysis by itself cannot determine which value of $z$ would be realized in a given physical situation, which must be determined empirically. We shall make further comments about this in the conclusions.

We end this Section by noting that we can recover the bound states
found in \cite{us2} from the poles of the $S$ matrix (\ref{25}) in the
positive imaginary axis. In the bound state sector $ m > E, $ by
setting $\; q = ip, \;\; p = \sqrt{m^{2} - E^{2}},\;$  as some arbitrary pole of the scattering matrix (\ref{25}), we obtain the condition determining the bound state
spectrum as
\begin{equation} \label{32}
   {\left ( \frac{m^{2} - E^{2}}{m^{2} + 1} \right )}^{ \nu}
   \frac{\Gamma (1 - 2\nu) \Gamma (1 + \nu - \frac{\alpha
   E}{\sqrt{m^{2} - E^{2}}})}{\Gamma (1 + 2\nu)\Gamma (1- \nu - \frac{\alpha  E}{\sqrt{m^{2} - E^{2}}})}
{\left ( \frac{j + \nu + \frac{\alpha (m + E)}{\sqrt{m^2 - E^2}}}{j - \nu +  \frac{\alpha (m + E)}{\sqrt{m^2 - E^2}}} \right )}
    = \frac{\xi_1 \cos(\theta_1 + \frac{z}{2})}
  {\xi_2 \cos(\theta_2 + \frac{z}{2})}.
\end{equation}
This equation determines the bound state energies in gapped graphene in the presence of a charge impurity, for the parameter range $\; 0 < \nu < \frac{1}{2} \; $ \cite{us2}.

\section{Conclusions}

In this paper we have analyzed the scattering sectors of the Hamiltonians for the gapless and gapped graphene in the presence of a charge impurity. We have shown that for a certain range of the system parameters, the corresponding Hamiltonians are not self-adjoint and therefore do not generate unitary time evolution with the usual boundary conditions. However, they can be made self-adjoint by a suitable choice of the boundary conditions, which requires the introduction of a real parameter in the problem. This self-adjoint extension parameter $z$ is a real number (defined modulo $2 \pi$ ) and for each value of $z$, we have an inequivalent quantization and spectral data for the system. In particular, the scattering state wave functions, the phase shifts and the $S$ matrices explicitly depend on the parameter $z$. This parameter however cannot be determined by theory alone, and has to be fixed empirically.

There are a variety of possibilities to fix this parameter empirically. First, the scattering matrix and the differential scattering cross-section determines the resistivity of the system \cite{kats}. A measurement of the resistivity would be a possible way to fix the parameter $z$. Second, it is well known that Friedel sum rule is affected by the self-adjoint extension \cite{moroz}, as it depends on the scattering phase shifts which are functions of the self-adjoint extension parameter. Thus the Friedel sum rule for graphene \cite{lin} would also depend on the parameter $z$. It follows that the integrated density of states would be $z$ dependent and the precise nature of this dependence which can be used to fix the value of $z$ is under investigation. 

Our analysis of the scattering sector also yields information about the bound states by looking at the poles of the $S$ matrix. Here we have recovered the results for the bound state spectrum which has already appeared elsewhere \cite{novi,us1,us2}. Measurement of local density of states (LDOS) using scanning tunneling spectroscopy can also yield information about the value of $z$ in a given graphene system. It should be noted that beyond the parameter range for which the self-adjoint extension is required, the established analyses of the scattering sector is unchanged. 

The analysis presented here can be extended to a variety of physical
effects in graphene. These include the analysis of the impact of generalized boundary conditions on screening effect in graphene \cite{screen, screen1, screen2, screen3, screen4} and on the study of impurities in bilayer graphene \cite{bi}. These are presently under investigation.


\vskip 1cm

\noindent
{\bf Acknowledgment}\\

This work was done within the framework of the Indo-Croatian Joint Programme of Cooperation in Science and Technology sponsored by the Department of Science and Technology, India (DST/INT/CROATIA/P-4/05) and the Ministry of Science, Education and Sports, Republic of Croatia. This work was supported by the Ministry of Science and Technology of the Republic of Croatia under contract No. 098-0000000-2865. 


\end{document}